\documentclass{article}

\usepackage[final]{neurips_2022}

\usepackage[utf8]{inputenc} 
\usepackage[T1]{fontenc}    
\usepackage{hyperref}       
\usepackage{url}            
\usepackage{booktabs}       
\usepackage{amsfonts}       
\usepackage{nicefrac}       
\usepackage{microtype}      
\usepackage{xcolor}         
\usepackage{graphicx}

\title{StyleGAN2-based Out-of-Distribution Detection for Medical Imaging}

\author{
    McKell Woodland\\
    MD Anderson Cancer Center\\
    Rice University\\
    \texttt{mewoodland@mdanderson.org} \\
    \And
    John Wood \\
    MD Anderson Cancer Center \\
    \And
    Caleb O'Connor \\
    MD Anderson Cancer Center \\
    \And
    Ankit B. Patel \\
    Baylor College of Medicine \\
    Rice University \\
    \And
    Kristy K. Brock \\
    MD Anderson Cancer Center
}

\begin{document}

\maketitle

\begin{abstract}

One barrier to the clinical deployment of deep learning-based models is the presence of images at runtime that lie far outside the training distribution of a given model.
We aim to detect these out-of-distribution (OOD) images with a generative adversarial network (GAN).
Our training dataset was comprised of 3,234 liver-containing computed tomography (CT) scans from 456 patients. 
Our OOD test data consisted of CT images of the brain, head and neck, lung, cervix, and abnormal livers.
A StyleGAN2-ADA architecture was employed to model the training distribution.
Images were reconstructed using backpropagation. 
Reconstructions were evaluated using the Wasserstein distance, mean squared error, and the structural similarity index measure.
OOD detection was evaluated with the area under the receiver operating characteristic curve (AUROC).
Our paradigm distinguished between liver and non-liver CT with greater than 90\% AUROC.
It was also completely unable to reconstruct liver artifacts, such as needles and ascites.

\end{abstract}

\section{Context}

Image segmentation is a critical task performed during radiotherapy to identify treatment targets and anatomical structures.
Manual segmentation is time-consuming [1].
It causes delays that have been correlated with lower survival rates [2], is subject to human variability [3], can lead to a lower quality of radiotherapy [4], and is incompatible with techniques that require frequent imaging to account for anatomical changes [5].
Automatic segmentation (autosegmentation) would not only be more efficient, 
but would also overcome intra-observer variability. 
Recently, deep learning algorithms have become the state-of-the-art autosegmentation models, with research spanning many anatomical regions and imaging modalities [6].
The liver, prostate, and spine are potentially the most accurately segmented structures and the most actively investigated [7].
Although deep learning-based liver segmentation models have been studied extensively, major challenges have diminished their clinical utility.
In this study, we focused on the challenge of poor model generalization.

The problem of poor generalization is exacerbated in the medical field due to limited data availability, especially of rare cases [8]. 
For instance, in Anderson et al., a deep learning-based liver segmentation model performed well on unseen test data of normal livers [9].
However, the same model completely failed on images where new information was presented (a stent and the presence of ascites).
We intend to mitigate the potential consequences of poor generalization by detecting when out-of-distribution (OOD) information is presented to the network.
The main contribution of our research is the detection of OOD images on which a trained segmentation model is likely to fail.
Secondarily, we show that the Wasserstein distance (WD) outperforms the mean squared error (MSE) and the structural similarity index metric (SSIM) as a reconstruction metric in OOD detection in the medical domain.

\section{Methods}
\subsection{Data}

Our training dataset contained 3,234 abdominal computed tomography (CT) scans from 456 patients at MD Anderson Cancer Center.
All axial slices that did not contain the liver were discarded.
To accentuate the liver, the data was windowed to a level 50 and a width 350, consistent with the preset values for viewing the liver in a commercial treatment planning system (RayStation v10, RaySearch Laboratories). 
Voxel values were mapped to the range [0, 255].
Each liver-containing axial slice was converted to a PNG image, resulting in 153,945 512x512 images.
1,000 of these images were used as an in-distribution test dataset.

Non-liver (brain, cervix, head and neck, and lung) datasets were constructed from 20 CTs.
Liver datasets with artifacts (needles and ascites) were constructed from 3 liver CTs with ascites and 51 liver CTs with needles.
All axial slices that did not contain the specified anatomical region/artifact were discarded.
Data was windowed to a level 50 and a width 350, voxel values were mapped to [0, 255], and slices were converted to PNGs.
Each of the four non-liver datasets consisted of 250 512x512 images, whereas the two abnormal liver datasets contained 150 images each.

\subsection{OOD detection}

A generative network was employed to model the distribution of training liver CT data.
Due to its state-of-the-art performance on high-resolution images, a StyleGAN2 network was selected as the generative model [10].
Specifically, the StyleGAN2 configuration of the official StyleGAN3 repository was utilized\footnote{\url{https://github.com/NVlabs/stylegan3}} [11].
The default parameters provided by the implementation were used, with the exception of changing $\beta_0$ to 0.9 in the Adam optimizer and disabling mixed precision.
This was done to stabilize training.
Transfer learning was employed by initializing training with weights from the Flickr-Faces-HQ dataset [12].
Data augmentation was performed by enabling mirroring (horizontal flipping) and adaptive discriminator augmentation [13].
For a computational ablation study demonstrating the benefits of these transfer learning and data augmentation choices, refer to [14].
The network was trained on a DGX with eight 40GB A100 GPUs, accessed using the XNAT platform [15].
It took approximately 4 days to complete the training (6,250 ticks with weights and metrics saved every 50 ticks).

All images were projected and subsequently reconstructed using the \emph{projector.py} file of the official StyleGAN2-ADA repository\footnote{\url{https://github.com/NVlabs/stylegan2-ada-pytorch}} with default parameters.
The network weights utilized were associated with the lowest Fr\'{e}chet Inception Distance achieved during training.
Three reconstruction scores were computed: the Wasserstein distance (WD) between intensity distributions, the mean squared error (MSE), and the structural similarity index measure (SSIM).
Only pixels within the human body were used in the WD and MSE calculations to minimize the effects of the surrounding black pixels.
The area under the receiver operating characteristic curve (AUROC) was calculated for each metric and dataset.
The liver dataset was randomly under-sampled to balance the classes for the AUROC calculation.

\section{Results}

\begin{figure}
    \centering
    \includegraphics[width=.75\linewidth]{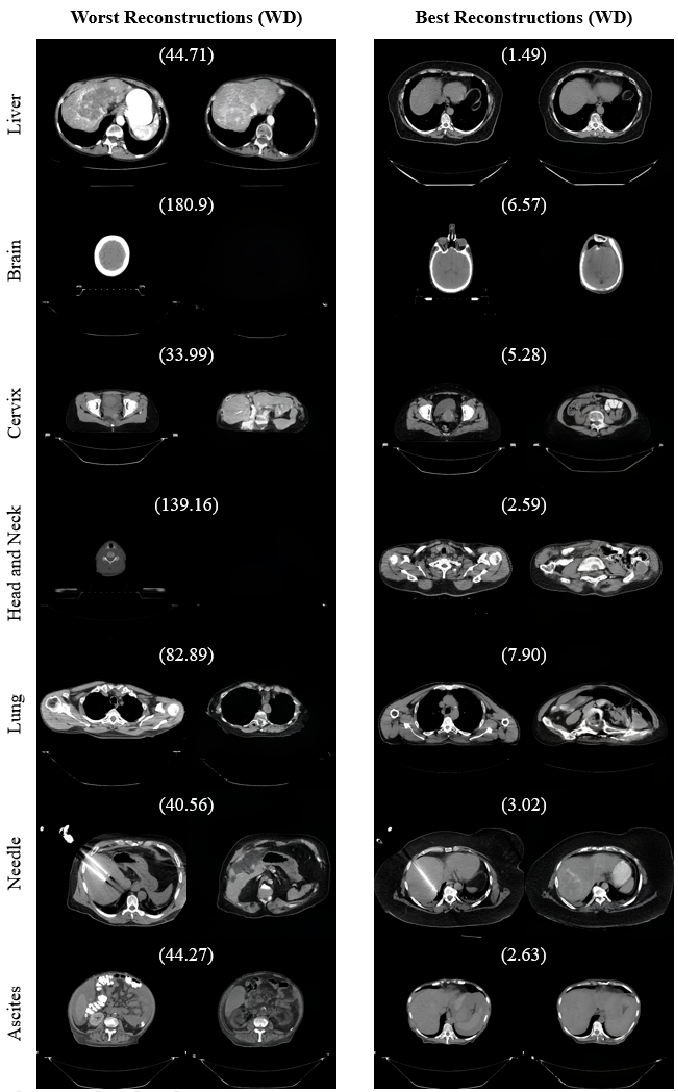}
    \caption{The best and worst reconstructions for each dataset (according the WD). 
    For each pair, the left image is the original and the right image is the reconstruction. 
    The number in parentheses is the WD between the pair.}
    \label{fig:recon}
\end{figure}

The generative adversarial network (GAN)-based OOD detection paradigm detected non-liver images with high AUROCs (Table \ref{tab:far-ood}).
The GAN seemed to model non-liver images by manipulating abdominal features (Figure \ref{fig:recon}), which resulted in large reconstruction errors.
The liver images that were classified as OOD often contained underrepresented artifacts.
For example, in the worst reconstruction of an in-distribution liver image in Figure \ref{fig:recon}, the network failed to model adjacent organs that had a high amount of contrast.

The SSIM demonstrated inconsistency across datasets (Table \ref{tab:far-ood}).
Because the WD outperformed the other metrics in terms of AUROC (except for one case, needles), we concluded that the WD was the best metric for OOD detection in medical images.

Our paradigm was also able to detect liver images with abnormalities (Table \ref{tab:far-ood}).
The GAN was completely unable to reconstruct needles and ascites (Figure \ref{fig:recon}).
This is a promising result because segmentation models often fail on images that contain such artifacts [9].
Because the GAN is unable to reconstruct uncommon attributes, we will be able to better detect these variations in future work.
We will be able to alert clinicians to these abnormalities, protecting against automation bias. 

\begin{table}
    \caption{The AUROCs for each OOD dataset and reconstruction metric.}
    \label{tab:far-ood}
    \centering
    \begin{tabular}{lccc}
        \toprule
        Dataset & WD-based AUROC & MSE-based AUROC & SSIM-based AUROC\\ 
        \midrule
        Brain & \textbf{.99} & .90 & .02\\
        Cervix & \textbf{.91} & .78 & .38 \\
        Head and Neck & \textbf{.97} & .96  & .05\\
        Lung & \textbf{.98} & .97 & .49\\
        Needles & .70 & .63 & \textbf{71}\\
        Ascites & \textbf{.66} & .50 & .16 \\
        \bottomrule
    \end{tabular}
\end{table}

\section*{Potential negative societal impact}

An OOD detector, deployed alongside an autosegmentation model, would have the benefit of warning clinicians when the model is likely to fail.
However, it may also have negative consequences.
For example, if the detector has a high false positive rate, it could contribute to alarm fatigue and create distrust.
On the other hand, it could lead to too much trust: if an autosegmentation model failed on an OOD input, and the OOD detector did not flag the input, the clinician may erroneously assume that the autosegmentation was correct.
Despite OOD detection showing promise in improving the safe deployment of autosegmentation models, the aforementioned risks should be carefully weighed by clinicians, computer scientists, and society as a whole.

\begin{ack}
This work was supported by the Tumor Measurement Initiative through the MD Anderson Strategic Iniative Development Program (STRIDE).
\end{ack}

\section*{References}

[1] Multi-Institutional. Human-computer interaction in radiotherapy target volume delineation: A prospective, multi-institutional comparison of user input devices. \emph{Journal of Digital Imaging}. 24(5), pp. 794-803 (2011). \url{https://doi.org/10.1007/s10278-010-9341-2}

[2] Chen, Z., King, W., Pearcey, R., Kerba, M., \& Mackillop, W.J. The relationship between waiting time for radiotherapy and clinical outcomes: A systematic review of the literature. \emph{Radiotherapy and Oncology}. 87(1), pp. 3-16 (2008). \url{https://doi.org/10.1016/j.radonc.2007.11.016}

[3] Nelms, B.E., Tom\'{e}, W.A., Robinson, G., \& Wheeler, J. Variations in the contouring of organs at risk: test case from a patient with oropharyngeal cancer. \emph{International Journal in Oncology Biology Physics}. 82(1), pp. 368-378 (2021). \url{https://doi.org/10.1016/j.ijrobp.2010.10.019}

[4] Saarnak, A.E., Boersma, M., van Bunningen, B.N.F.M., Wolterink, R., \& Steggerda, M.J. Inter-observer variation in delineation of bladder and rectum contours for brachytherapy of cervical cancer. \emph{Radiotherapy and Oncology}. 56(1), pp. 37-42 (2000). \url{https://doi.org/10.1016/S0167-8140(00)00185-7}

[5] Sheng, K. Artificial intelligence in radiotherapy: a technological review. \emph{Frontiers of Medicine}. 14(4), pp. 431 (2020). \url{https://doi.org/10.1007/s11684-020-0761-1}

[6] Cardenas, C.E., Yang, J., Anderson, B.M., Court, L.E., \& Brock, K.B. Advances in Auto-Segmentation. \emph{Seminars in Radiation Oncology}. 29(3), pp. 185-197 (2019). \url{https://doi.org/10.1016/j.semradonc.2019.02.001}

[7] Zhou, S.K., et al. A review of deep learning in medical imaging: Image traits, technology trends, case studies with progress highlights, and future promises. \emph{Proceedings of the IEEE}. 109(5), pp. 820-838 (2021). \url{https://doi.org/10.1109/JPROC.2021.3054390}



[8] Baydargil, H.B., Park, J.-S., \& Kang, D.-Y. Anomaly analysis of alzheimer's disease in PET images using an unsupervised adversarial deep learning model. \emph{Applied Sciences}. 11(5) (2021). \url{https://doi.org/10.3390/app11052187}

[9] Anderson, B.M., et al. Automated contouring of contrast and noncontrast computed tomography liver images with fully convolutional networks. \emph{Adv Radiat Oncol}. 6, pp. 100464 (2021). \url{https://doi.org/10.1016/j.adro.2020.04.023}



[10] Karras, T., Laine, S., Aittala, M., Hellsten, J., Lehtinen, \& J., Aila, T. Analyzing and improving the images quality of StyleGAN. In: CVPR 2020. pp. 8107-8116. IEEE (2020). \url{https://doi.org/10.1109/CVPR42600.2020.00813}

[11] Karras, T., et al. Alias-free generative adversarial networks. In: Ranzato, M., Beygelzimer, A., Dauphin, Y., Liang, P., Vaughan, J.W. (eds.) NeurIPS 2021. vol. 34, pp. 852-863. Curran Associates, Inc. (2021)

[12] Karras, T., Laine, S., \& Aila, T. A style-based generator architecture for generative adversarial networks. In: CVPR 2019. pp. 4396-4405. IEEE (2019). \url{https://doi.org/10.1109.CVPR.2019.00453}

[13] Karras, T., Aittala, M., Hellsten, J., Laine, S., Lehtinen, J., \& Aila, T. Training generative adversarial networks with limited data. In: Larochelle, H., Ranzato, M., Hadsell, R., Balcan, M., Lin, H. (eds.) NeurIPS 2020. vol 33, pp. 12104-12114. Curran Associates, Inc. (2020)

[14] Woodland, M. et al. Evaluating the Performance of StyleGAN2-ADA on Medical Images. In: Zhao, C., Svoboda, D., Wolterink, J.M., Escobar, M. (eds) Simulation and Synthesis in Medical Imaging. SASHIMI 2022. Lecture Notes in Computer Science, vol 13570. Springer, Cham. \url{https://doi.org/10.1007/978-3-031-16980-9\_14}

[15] Marcus, D.S., Olsen, T.R., Ramaratnam, M., \& Buckner, R.L. The extensible neuroimaging archive toolkit: an informatic platform for managing, exploring, and sharing neuroimaging data. \emph{Neuroinformatics}. 5, pp. 11-33 (2007). \url{https://doi.org/10.1385/ni:5:1:11}
\end{document}